\documentclass[reprint]{aastex63}

\usepackage{subfigure}
\usepackage{float}

\begin{document}

\title{A unified model of solar prominence formation}

\author{C. J. Huang}
\affiliation{School of Astronomy \& Space Science, Nanjing University, Nanjing 210023, China}
\affiliation{Key Lab of Modern Astron. \& Astrophys. (Ministry of Education), Nanjing University, China}

\author[0000-0002-4205-5566]{J. H. Guo}
\affiliation{School of Astronomy \& Space Science, Nanjing University, Nanjing 210023, China}

\author[0000-0002-9908-291X]{Y. W. Ni}
\affiliation{School of Astronomy \& Space Science, Nanjing University, Nanjing 210023, China}

\author{A. A. Xu}
\affil{State Key Laboratory of Lunar and Planetary Sciences, Macau University of Science and Technology, Macau, China}

\author[0000-0002-7289-642X]{P. F. Chen}
\affiliation{School of Astronomy \& Space Science, Nanjing University, Nanjing 210023, China}
\affiliation{Key Lab of Modern Astron. \& Astrophys. (Ministry of Education), Nanjing University, China}

\correspondingauthor{P. F. Chen}
\email{chenpf@nju.edu.cn}

\begin{abstract}
Several mechanisms have been proposed to account for the formation of solar prominences or filaments, among which direct injection and evaporation-condensation models are the two most popular ones. In the direct injection model, cold plasma is ejected from the chromosphere into the corona along magnetic field lines; In the evaporation-condensation model, the cold chromospheric plasma is heated to over a million degrees and is evaporated into the corona, where the accumulated plasma finally reaches thermal instability or non-equilibrium so as to condensate to cold prominences. In this paper, we try to unify the two mechanisms: The essence of filament formation is the localized heating in the chromosphere. If the heating happens in the lower chromosphere, the enhanced gas pressure pushes the cold plasma in the upper chromosphere to move up to the corona, such a process is manifested as the direct injection model. If the heating happens in the upper chromosphere, the local plasma is heated to million degrees, and is evaporated into the corona. Later, the plasma condensates to form a prominence. Such a process is manifested as the evaporation-condensation model. With radiative hydrodynamic simulations we confirmed that the two widely accepted formation mechanisms of solar prominences can really be unified in such a single framework. A particular case is also found where both injection and evaporation-condensation processes occur together.
\end{abstract}


\section{Introduction} \label{sec:intro}

Solar filaments are cool and dense plasma suspended in the hot and tenuous corona above the photospheric magnetic neutral lines. They appear dark against the solar disk typically observed in H$\alpha$, but look bright above the solar limb, where they are called solar prominences. Solar filaments are an intriguing phenomenon in the solar atmosphere for several reasons \citep{chen20}: Their formation is often related to thermal instability and/or thermal non-equilibrium \citep{anti99, klim19, antolin20}, which is an opposite process of coronal heating; their evolution is strongly related to the mass circulation among different layers of the solar atmosphere; their oscillations can be applied to decipher the otherwise elusive magnetic field in the corona; their eruptions are intimately related to the two major solar eruptions, i.e., solar flares and coronal mass ejections (CMEs).

The typical mass of a solar filament is on the order of $10^{14}$--$10^{15}$ g \citep{pare14}, with a density about two orders of magnitude larger than the ambient corona. Therefore, one important issue is how the cold dense structure is formed. The formation of solar filaments has been an interesting topic for decades, and three mechanisms have been proposed \citep{mack10}, i.e., chromospheric evaporation plus coronal condensation model (or called condensation model for short), direct injection model, and levitation model. The levitation model claims that magnetic flux tubes emerge from the solar subsurface to the atmosphere, where cold chromospheric plasma is levitated to the corona straight above the magnetic neutral line \citep{rust94, lite05, okam08}. In accord with the fact that there were only a few reports favoring this model, we tend to think that what emerging magnetic flux produces is arch filament systems, not solar filaments. In contrast, there is much more evidence supporting the formation of filaments via coronal condensation \citep{berg12, liu12, Vial20} and direct injection \citep{chae00, wang18, yan20}. In particular, the sudden appearance of H$\alpha$ threads in filaments from nowhere can only be explained by the condensation model. Therefore, these two mechanisms are the most popular models, and have been extensively simulated via numerical simulations \citep{anch88, wust90, anti91, lee95, karp01, xia11}. In the evaporation plus condensation model, cold chromospheric plasma is heated locally to high temperatures (up to millions of degrees) and then is evaporated to the corona, where accumulated hot plasma suffers from thermal instability or thermal non-equilibrium, and then cools down to form cold filaments. In the direct injection model, cold chromospheric plasma is somehow driven to be ejected to the corona, and the cold plasma is either trapped in the magnetic dips in the corona to form a quasi-static filament or drains down along the other side of the magnetic flux tube to form a dynamic filament. The difference between the two models is that heated plasma is injected from the chromosphere to the corona in the condensation model and cold plasma is injected from the chromosphere to the corona in the injection model.

While both the condensation model and the direct injection model well explain the formation of many solar filaments, one important issue that has not yet been answered in the modeling is what primary process is responsible for the two models, i.e., what drives the localized heating in the condensation model and the chromospheric upflow in the injection model? The most probable answer is magnetic reconnection in the chromosphere, as evidenced by observations \citep{chae00, yan16, zoup16, yang19}. However, a further question can be asked: What leads to the entirely different behaviors in the two models if both are driven by magnetic reconnection in the low atmosphere? Here we propose a conjecture in order to unify the two models, i.e., while both models are due to localized heating in the chromosphere, if the heating is situated higher in the upper chromosphere, the plasma in the upper chromosphere would be heated to millions of Kelvin so that it is evaporated to the corona, which later condenses to form a cold filament when thermal instability happens or thermal non-equilibrium happens; If the heating is situated in the lower chromosphere, the plasma in the lower chromosphere is heated locally, whose enhanced gas pressure pushes the cold plasma in the upper chromosphere to be ejected to the corona.
In order to confirm whether the unified model works, in this paper we perform a series of numerical simulations, where localized heating is specified at different layers of the solar chromosphere. This paper is organized as follows. The numerical simulation setup is introduced in \S\ref{sec2}, and the numerical results are described in \S\ref{sec3}, which is followed by discussions in \S\ref{sec4}.

\section{Numerical setup}\label{sec2}
\subsection{Governing equations}
To simulate the formation of solar filaments via either the condensation model or the injection model, we solve the one-dimensional (1D) hydrodynamic equations, with fully ionized plasma restrained in a single dipped magnetic flux tube. Both thermal conduction and radiative cooling are taken into account. Gravity is downward in the vertical direction and only the magnetic field-aligned component is really effective. Following \citet{xia11}, we use the Message Passing
Interface Adaptive Mesh Refinement Versatile Advection Code \citep[MPI-AMRVAC,][]{kepp03, xia18} to solve the following 1D hydrodynamic equations:

\begin{equation}
\begin{array}{cc}
     & \frac{\partial\rho}{\partial t} + \frac{\partial}{\partial s}(\rho v) = 0,  \\
     & \frac{\partial}{\partial t}(\rho v) + \frac{\partial}{\partial s}(\rho v^{2}+p) = \rho g_{||}(s), \\
     & \frac{\partial\varepsilon}{\partial t} +\frac{\partial}{\partial s}(\varepsilon v + pv) = \rho g_{\vert\vert}(s)v +H(s,t) - n_{\rm{H}}n_{\rm{e}}\Lambda(T) + \frac{\partial}{\partial s}(\kappa\frac{\partial T}{\partial s}),
\end{array}
\end{equation}
\noindent
where $g_{\vert\vert}$ is the component of gravity along the magnetic field line, which is also the axis of the arc coordinate $s$, $\rho$ is the mass density, $v$ is the velocity along the coordinate axis; $\varepsilon=\rho v^2/2+p/(\gamma -1)$ is the total energy density, including the kinetic and internal energies. The plasma is approximated to be fully ionized. Therefore, considering the abundance of helium, we have $\rho=1.4 m_{\rm H} n_{\rm H}$, and $p=2.3n_{\rm H} k_{\rm B} T$ , where $k_{\rm B}$ is the Boltzmann constant; $m_{\rm H}$ and $n_{\rm H}$ are the proton mass and number density, respectively. Besides, $\gamma=5/3$ is the adiabatic index, $\kappa = 10^{-6} T^{5/2} \rm{erg~cm^{-1}~{s^{-1}}~K^{-1}}$ is the Spitzer heat conductivity. The volumetric heating rate $H(s,t)$ consists of a background heating $H_b(s)$ and a localized heating $H_l(s,t)$, where $H_b(s)$ is a steady heating which exists during the whole simulation in order to maintain the hot corona. Although its nature remains elusive, its distribution is commonly speculated to decay with height. Following the previous works \citep[e.g.,][]{anti99}, we set

\begin{equation}
    H_{b}(s)=\left\{
    \begin{array}{cc}
         E_{0}\exp(-s/H_s),     & s<L/2;  \\
         E_{0}\exp[-(L-s)/H_s],   & 2/L \leq s <L,
    \end{array}\right.
\end{equation}
\noindent
where $E_0=1.0\times10^{-4}\ {\rm erg~cm}^{-3}~{\rm s}^{-1}$ and the scale height $H_s$ is half of the full length of the magnetic field line $L$. Before the localized heating is imposed, the simulated loop should reach a hydrostatic equilibrium resembling the quiet solar atmosphere under the effect of the background heating.

As for the cooling due to optically thin radiation, we adopt the radiative loss function $\Lambda(T)$ obtained by \citet{colg08}. In the low temperature regime $T<1.6\times 10^4$ K, the radiative cooling function vanishes. A detailed description of the radiative loss function can be found in \citet{xia11}.

\subsection{Initial and boundary conditions}

\begin{figure}
      \centering
      \includegraphics[width=0.60\textwidth]{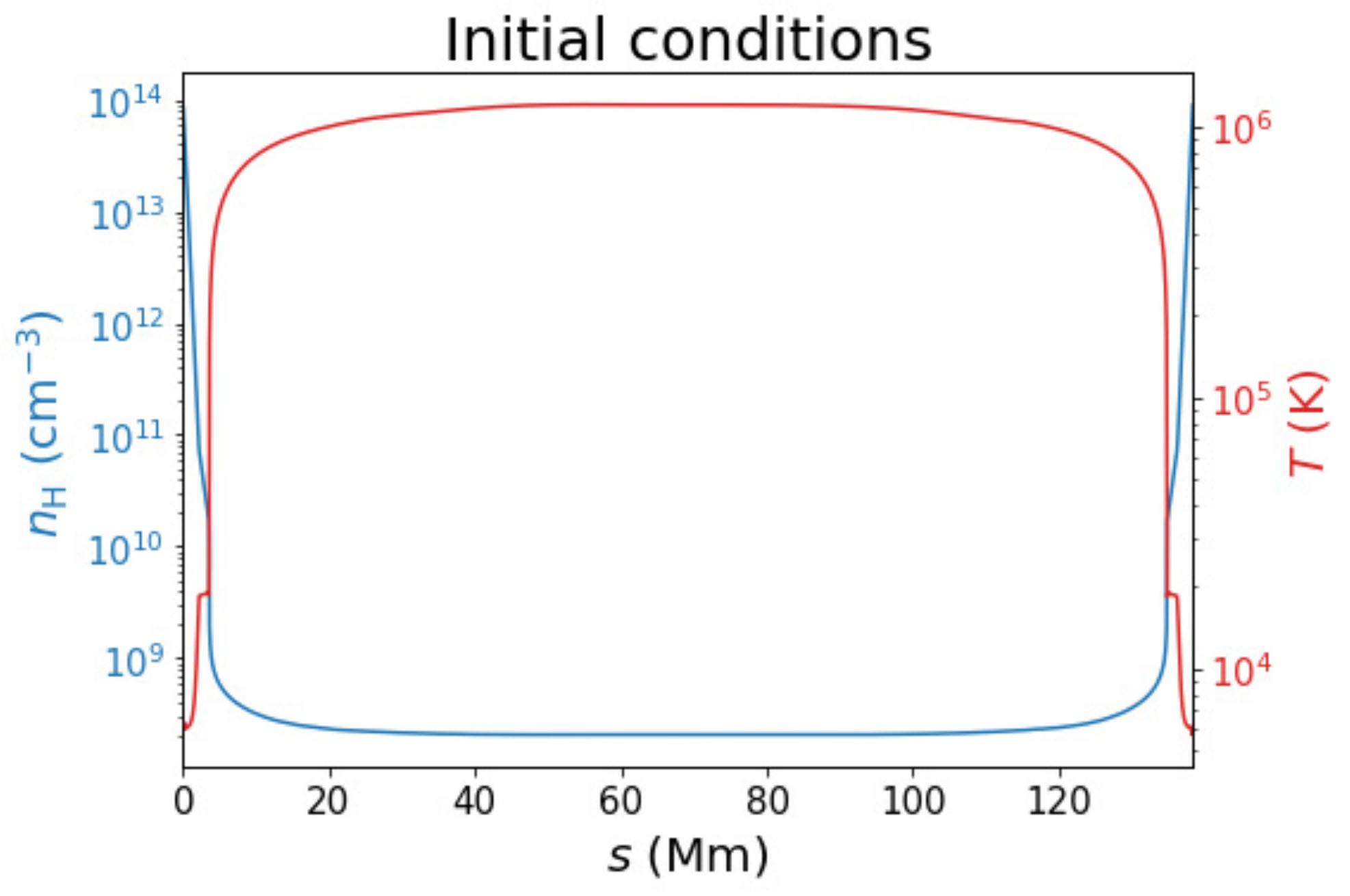}
      \caption{Temperature and density distributions in the initial conditions of our simulations, which are obtained via 9000 s relaxation. }
	\label{fig1}
 \end{figure}

Similar to \citet{xia11}, the rigid magnetic flux tube is composed of two vertical legs, two shoulders which are described as two quarter-circles, and a sinusoidal-shaped magnetic dip connecting the shoulders. The gravity distribution on the left half of such a geometry is described as follows (note that the right half is simply symmetric about the midpoint):

\begin{equation}
    g_{||}(s)=\left\{
    \begin{array}{cc}
         -g_{\sun},     & s<s_1;  \\
         -g_{\sun}\cos\frac{s-s_{1}}{2(s_{2}-s_1)}\pi,  & s_{1} < s \le s_{2}; \\
         g_{\sun}\frac{\pi d}{2(L/2-s_2)}\sin\frac{s-s_2}{L/2-s_2}\pi,  & s_{2} < s \le L/2,
    \end{array}\right.
	\label{eq3}
\end{equation}
\noindent
where $g_{\sun}=2.7\times 10^4$ m s$^{-2}$ is the gravity near the solar surface, we set $s_1=4.0$ Mm as the length of each loop leg. The radius of the quarter-circles is $r=8.0$ Mm, so $s_{2}=s_{1}+\pi r/2= 16.56$ Mm. $D=4$ Mm is the depth of the magnetic dip, which leads to the length of the sinusoidal dip being 52.63 Mm. All together, the full length of the magnetic loop is 138.38 Mm. In order to construct stable initial thermal conditions, we initially set a hydrostatic equilibrium with the temperature distribution similar to \citet{Zhou17} described as

\begin{equation}
    T(z)=T_{1}+\frac{1}{2}(T_{2}-T_{1})\tanh(\frac{z-h_{0}}{w_{0}}+1),
\end{equation}
\noindent
where $T_{1}=6000$ K and $T_2=10^6$ K are the plasma temperatures at the footpoint and the corona, respectively, $h_0=2.5$ Mm is the height of transition region between the chromosphere and the corona, and $w_0=0.25$ Mm. Density is calculated in order to balance the gravity. Such a state is in force equilibrium, but not in thermal equilibrium. Once the background heating is introduced in the energy equation, the whole loop evolves. After $t$=9000 s, the loop reaches a quasi-stationary state under the influence of gravity and the background heating, and such a state is utilized as the initial conditions for our later simulations. Figure \ref{fig1} displays the temperature distribution as the red line and the density distribution as the blue line. It is seen that the temperature is the highest at the midpoint with $T_{max}=1.16\times 10^6$ K and the density is the lowest with $n_H=2.05\times 10^8$ cm$^{-3}$. It is also found that the height of the transition region is about 3.7 Mm after relaxation. During the whole evolution of our later simulations, the boundary conditions at the two footpoints are fixed, which means that the density and temperature are unchanged and the velocity is set to be zero.

\subsection{Localized heating}
The height of the localized heating in the chromosphere is the key to our unified model. It is reasonable to investigate the effect of the heating height since it was already confirmed that magnetic reconnection at different heights in the chromosphere can produce different observational characteristics \citep{shib96, chen01, jian12}. For simplicity, in this paper the heating function has a Gaussian shape both in space and in time:

\begin{equation}
    H_l(s,t)=\left\{
    \begin{array}{lc}
        H_{peak}\exp{[-(s-s_{peak})^2/s^2_c]}\exp{[-(t-t_{peak})^2/t^2_{c1}]}, &  t<t_{peak}; \\
        H_{peak}\exp{[-(s-s_{peak})^2/s^2_c]} \exp{[-(t-t_{peak})^2/t^2_{c2}]},  &  t \ge t_{peak}. \\
    \end{array}\right.
	\label{eq5}
\end{equation}

Table \ref{tab1} lists the parameters in the localized heating function. According to our conjecture, the heating is expected to be at a higher altitude for the evaporation plus condensation model (cases denoted by ECO*), where $s_{peak}$ is assumed to be 3.59 Mm, and at a lower altitude for the injection model (cases denoted by INJ*), where  $s_{peak}$ is assumed to be 1.91 Mm. Considering that the transition region is at $s=3.7$ Mm in our numerical model, the heating height in cases ECO* is 0.11 Mm below the transition region, i.e., upper chromosphere, whereas the heating height in cases INJ* is 1.68 Mm below the transition region, i.e., the lower chromosphere. Besides, considering that both the magnetic field and the released energy decrease with height, we use weaker heating amplitude for the upper chromosphere and stronger heating amplitude for the lower chromosphere. Initially the simulation domain is discretized into 1500 grid points with a 5-level grid mesh refinement, which results in a maximum spatial resolution of 2.88 km.

\begin{deluxetable*}{lcccccl}\label{tab1}
\tablenum{1}
\tablecaption{Parameters of the localized heating function in 6 cases.}
\tablewidth{0pt}
\tablehead{
\colhead{Case} & \colhead{$H_{peak}$} &  \colhead{$s_{peak}$}  & \colhead{$s_c$}  & \colhead{$t_{c1}$} & \colhead{$t_{c2}$} & \colhead{Comments}\\
\colhead{} &\colhead{(${\rm erg~cm}^{-3}~{\rm s}^{-1}$)}& \colhead{(Mm)}  & \colhead{(Mm)}  & \colhead{(s)} & \colhead{(s)} &\colhead{}
}
\startdata
ECO1 & 1.0 & 3.59 & 0.15 & 100 & 333 & {Standard case for the evaporation-condensation model} \\
ECO2 & 1.0 & 3.59 & 0.15 & 100 & 133 & {No filament is formed} \\
INJ1 & 3.0 & 1.91 & 0.15 & 100 & 133 & {Standard case for the injection model} \\
INJ2 & 3.0 & 1.91 & 0.15 & 133 & 133 & {No cold upflow is formed} \\
INJ3 & 3.0 & 1.91 & 0.15 & 100 &  67 & {Cold flow rises and drains back } \\
INJ4 & 3.0 & 1.91 & 0.15 & 100 & 167 & {Cold upflow overshoots to the other side} \\
\enddata
\end{deluxetable*}

\section{Results} \label{sec3}

\begin{figure*}
\centering
 \includegraphics[width=0.80\textwidth]{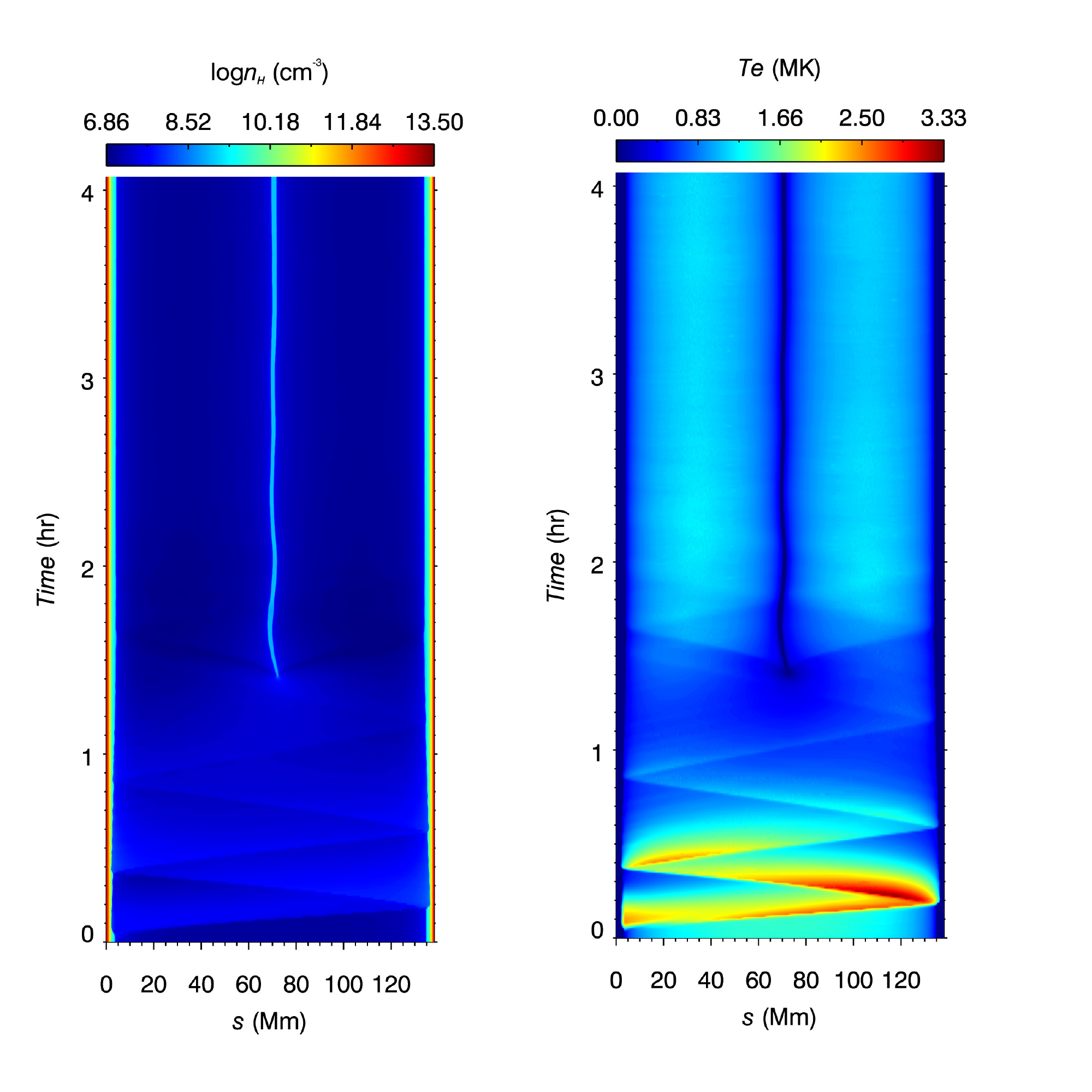}
	\caption{Evolutions of the plasma density distribution (left panel) and temperature distribution (right panel) in case ECO1, which is simulated to reproduce the evaporation--condensation mechanism for filament formation. The horizontal axis is the spatial coordinate along the magnetic loop, and the vertical axis is time.}
	\label{fig2}
\end{figure*}

As listed in Table \ref{tab1}, we perform simulations in 6 cases. In cases ECO1--2, the localized heating is assumed to be at the upper chromosphere, and in cases INJ1--4, the localized heating is at the lower chromosphere. Figure \ref{fig2} displays the evolutions of the density distribution (left panel) and the temperature distribution (right panel) for the case ECO1, which is the standard case for the evaporation-condensation model. It is seen that as the localized heating is introduced in the upper chromosphere of the left-sided footpoint, plasma is heated to evaporate into the corona. As a result, both the density and temperature increase in the coronal part of the magnetic loop. When the heating reaches its maximum, strong evaporation is produced. The evaporation upflow generates a shock wave propagating along the magnetic loop with a speed of 234.6 km s$^{-1}$ as evidenced by the red and yellow ridges in the right panel of Figure \ref{fig2}. When the shock wave reaches the right footpoint of the magnetic loop, it bounces back. In the following period from $t=0.2$ hr to $t=1.5$ hr, the shock wave bounces back and forth between the two footpoints of the loop, slowing down gradually. As the evaporating plasma moves up, part of the material goes down to the right-handed footpoint, and part of the material accumulates in the corona, which would enhance thermal radiation. It is seen from the right panel of Figure \ref{fig2} that the plasma temperature near the midpoint of the magnetic loop, i.e., $s\sim$69 Mm, continues to decreases from $t=0.46$ to $t=1.30$ hr except when the shock wave passes through. At around $t=1.4$ hr, the plasma density increases and the temperature decreases drastically at $s\sim 72.0$ Mm, as indicated by the light blue segment in the left panel and the dark segment in the right panel. That is to say, a filament thread is formed. After tens of minutes of oscillations caused by the interplay among gravity, gas pressure, and the bouncing shock, the filament thread grows in length, which saturates at 2.0 Mm, located 1.11 Mm to the right of  the midpoint of the magnetic dip.

It is expected that whether the thermal instability occurs or not depends on the density of the accumulated plasma in the corona. In case ECO2, the localized heating has the same amplitude and the rising time as case ECO1, but with a shorter declining time. Therefore, the total energy deposited into the upper chromosphere is reduced. As a consequence, no thermal instability occurs, and no filament thread is formed.

\begin{figure*}
\centering
 \includegraphics[width=0.80\textwidth]{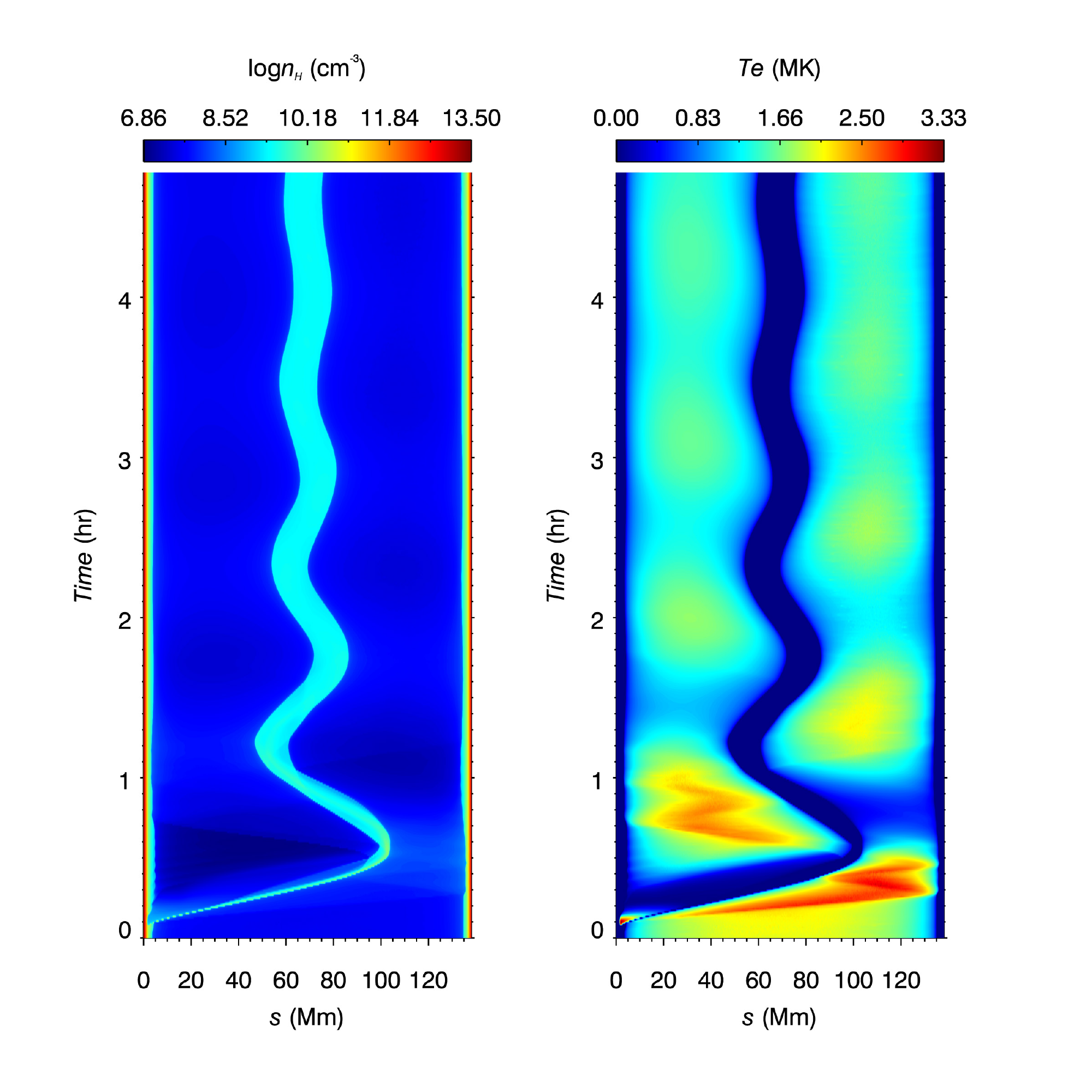}
	\caption{Evolutions of the plasma density distribution (left panel) and temperature distribution (right panel) in case INJ1, which is simulated to reproduce the direct injection mechanism for filament formation. The horizontal axis is the spatial coordinate along the magnetic loop, and the vertical axis is time.}
	\label{fig3}
\end{figure*}

In order to reproduce the direct injection mechanism, we simulate another 4 cases denoted as cases INJ1--4 where the impulsive heating is localized in the lower chromosphere at an altitude of 1.91 Mm. One successful scenario is case INJ1, where the spatial extent and the rising time of the heating profile are the same as case ECO1, but the declining time is reduced to 133 s. The evolution of the density and temperature distributions is displayed in Figure \ref{fig3}. It is seen that after heat is deposited in the lower chromosphere of the left-sided footpoint, the local plasma is heated to high temperatures (the yellow and red tiny area near the left footpoint). The enhanced gas pressure in the lower chromosphere pushes the upper cold dense plasma to rise up to the corona with a speed of 68.4 km s$^{-1}$, as indicated by the light blue segment in the left panel and the dark blue segment in the right panel. The sudden rise motion of the cold plasma is like a piston, driving a weak shock wave ahead of the cold material, manifested as the red ridge with a propagating velocity of 138.2 km s$^{-1}$. When the shock wave reaches the right-sided footpoint, it bounces back and collides with the rightward cold mass flow to the left of the midpoint of the magnetic dip. Therefore, it is seen that in this case the blue-coded cold material can not overshoot the right shoulder of the magnetic dip mainly due to the rebounding shock wave. Later, the directly formed filament thread oscillates around the midpoint of the magnetic dip under the influence of the field-aligned gravity. The filament thread in case INJ1 has a length of 15.4 Mm.

The characteristics of the filament formation and its dynamics strongly depend on the properties of the heating process. In cases INJ2--INJ4, we change the spatial extent and the temporal profile, and find that (1) if the heating in the lower chromosphere is more gradual in the rising phase as in case INJ2, heat conduction plays a role, and even the upper chromosphere is heated. As a result, no cold upflow is produced; (2) if the heating decays too quickly as in case INJ3, the enhanced gas pressure cannot accelerate the cold upflow to a sufficiently high speed. As a result, the cold upflow drains back to the left-sided footpoint; (3) if the heating decays too slowly in time as in case INJ4, the enhanced gas pressure continues to push the cold upflow, so that the formed filament thread overshoots over the shoulder on the right side and drains down to the right-sided footpoint, forming a dynamic filament.

\section{Discussion}\label{sec4}

\subsection{A unified model of filament formation}
For two reasons, the material of many solar filaments suspended in the corona mainly originates from the chromosphere, rather than the ambient corona. First, the mass of a filament is often more than the local corona can provide \citep{vial15}. Second, the element abundance in solar filaments is closer to that in the chromosphere, rather than that in the ambient corona \citep{spicer98, song17}. The mass supply from the chromosphere to the corona can be realized in two completely different ways. One is the direct injection of the cold chromospheric plasma into the corona, and the other is that the cold chromospheric plasma is heated to evaporate into the corona, where it cools down via either thermal instability \citep{park53} or thermal non-equilibrium \citep{klim19}. As a result, the direct injection and evaporation--condensation models are the two most popular mechanisms for the formation of solar filaments. In the former model, cold chromospheric material is heated first, presumably via magnetic reconnection. In the latter model, cold chromospheric material is pushed to the corona, which was also proposed to be triggered by magnetic reconnection during magnetic flux submergence \citep{wang99}. In the past decades, the two models were treated separately. However, it seems that both mechanisms are related to magnetic reconnection in the solar low atmosphere. This arises a question: What is the fundamental difference between the two mechanisms?

In this paper, we conjectured that it the height of magnetic reconnection that distinguishes the two apparently different models, i.e., if magnetic reconnection happens locally in the upper chromosphere, the in-situ plasma is heated to evaporate into the corona, fitting into the evaporation--condensation model; if magnetic reconnection happens in the lower chromosphere, the in-situ plasma is heated and the enhanced gas pressure pushes the upper chromosphere to rise into the corona directly, fitting into the direct injection model. We tested this idea by 1D hydrodynamic simulations, where the effect of magnetic energy release is represented by in-situ heating for simplicity as done in previous 1D simulations \citep{anti91, karp01, luna12}. According to our simulations, indeed we found that (1) if the localized heating happens in the upper chromosphere, the in-situ cold plasma is heated and evaporated into the corona, where it condenses to form a filament thread via thermal instability; (2) if the localized heating happens in the lower chromosphere, the evolution fits the direct injection model very well. Depending on the released energy, the surge-like upflows have different rising speeds. As a result, the cold material is either trapped in the coronal magnetic dip or keeps moving until draining down to the other footpoint of the magnetic loop.

\begin{figure*}
\centering
 \includegraphics[width=0.80\textwidth]{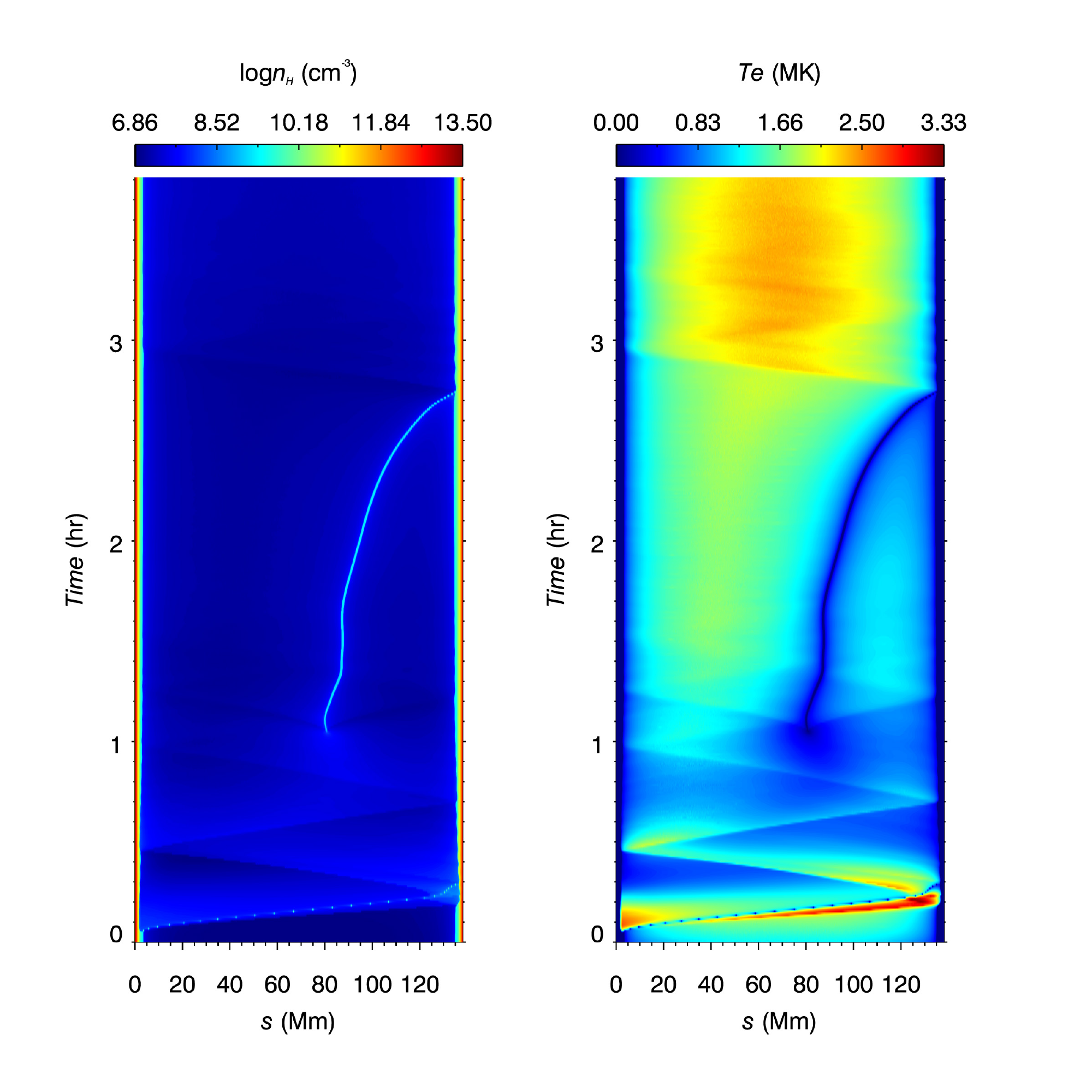}
	\caption{Evolutions of the plasma density distribution (left panel) and temperature distribution (right panel) when the localized heating is assumed to be in the middle chromosphere. The horizontal axis is the spatial coordinate along the magnetic loop, and the vertical axis is time.}
	\label{fig4}
\end{figure*}

The energy deposit in the upper chromosphere and the lower chromosphere is two extremes, which well fit into the evaporation--condensation model and direct injection model, respectively. Since reconnection may happen at any heights in the solar chromosphere \citep{jian12}, it might be expected to see that energy deposit at certain heights may lead to compound features of both the evaporation--condensation model and the direct injection model. For this purpose, we perform another numerical simulation, where the localized heating is put at $s=2.56$ Mm, with its $H_{peak}$ the same as case INJ1 and other parameters the same as case ECO1. Figure \ref{fig4} displays the evolution of the density distribution in the left panel and temperature distribution in the right panel. It is revealed that at $t=0.06$ hr, surge-like cold plasma is ejected from the left-sided footpoint to the corona as evidenced by the light blue ridge in the left panel and the dark blue ridge in the right panel, forming a dynamic filament thread first. A shock wave is also excited ahead of the moving filament thread as indicated by the red front. The speed of the filament thread is so large that the cold material overshoots the apex of the magnetic dip and drains down to the right-sided footpoint at $t=0.28$ hr. When both the shock wave and the cold thread hit the right-sided footpoint, reflective shock waves are generated, which propagate in the magnetic loop, bouncing back and forth between the two footpoints of the magnetic loop. At the same time, part of the heated chromospheric material remains in the corona. At $t=1.00$ hr, thermal instability is triggered to the right of the midpoint of the magnetic loop, and a second filament thread is formed at $s=81.2$ Mm. This filament thread sustains for 1.72 hr before draining down to the right-sided footpoint. Therefore, in this case filament threads form in a repeated way, one via the direct injection mechanism, and the other via the evaporation--condensation mechanism. Since the solar photosphere and chromosphere are turbulent, small-scale magnetic energy release and deposit can happen randomly either in the horizontal plane or in the height direction. Not only is the random deposit of energy responsible for the formation of discrete threads in a filament body \citep{zhou20}, but also the random deposit of energy at different heights produces either evaporation--condensation or direct injection processes for the filament formation. All these random processes sum up to compose the dynamic life of any solar filament, as elucidated by \citet{liu12}.

It is noted that the magnetic dip in all the cases listed in Table \ref{tab1} has an altitude of 8 Mm, which is in the typical height range of active-region filaments \citep{mack10}. In order to check whether the injection model works well for taller quiescent filaments, which have heights up to more than 40 Mm, we perform other simulations, where the height of the magnetic dip is uplifted to 20, 30, and 40 Mm by increasing the length of the vertical loop leg, i.e., $s_1$ in Equation (\ref{eq3}). It is found that if the magnetic dip is as high as 20 Mm, the same parameters in case INJ1 can still guarantee the formation of a filament via direct injection. However, if the magnetic dip is uplifted to 30 Mm, the parameters in case INJ1 fail to make filament formation via direct injection. The localized heating center has to be shifted from $s_{peak}$=1.91 Mm to 2.2 Mm, so that a filament thread with a length of 8.4 Mm is formed via direct injection. This is understood as follows: In order to eject cold materials to a higher altitude, the velocity has to be larger. With the same impulse, it is easier for lighter mass to be accelerated to larger velocities. When the magnetic dip is uplifted to 40 Mm, a stable filament thread cannot be formed by simply shifting up the heating center in the direct injection model. However, we find that by prolonging the decay phase of the localized heating, i.e., $t_{c2}$ in Equation (\ref{eq5}), a filament thread can be formed as well. As the localized heating is shifted higher and higher, the direct injection model naturally transforms to the evaporation--condensation model.

Besides, recent simulations and observations indicated that the formation of some filaments might be due to thermal instability after reconnection in the low corona \citep{kane17, zhao17, li19}. In this mechanism, no chromospheric evaporation is involved. Hence, it is expected that the element abundance in these often small filaments might be similar to that of the corona, rather than the solar chromosphere \citep{chen20}.

\subsection{How is mass supplied to the corona}

According to \citet{karp08}, if the localized heating is repetitive rather than steady, then filament threads can be formed only when the interval between the successive heating is less than the coronal radiative cooling time. However, in our case ECO1, a filament thread is formed even if there is a single heating pulse. The apparent discrepancy is due to that the heating amplitude in our case is 10 times stronger than in their case. We also perform other simulations with weaker heating amplitude, and it is found that when $H_{peak}$ is reduced to 0.1 erg cm$^{-3}$ s$^{-1}$, no filament thread is formed, consistent with \citet{karp08}. If $H_{peak}$ is set to be 0.3 erg cm$^{-3}$ s$^{-1}$, a small segment of filament thread can be formed, but soon it is swept by the reflected shock wave, and drains down to the left-sided footpoint.

Nevertheless, our simulations indeed tell us that given strong amplitudes, even a single heating pulse can also induce the formation of a filament thread, once the deposited energy is sufficiently strong. In this case, a major part of the accumulated plasma in the corona comes from the evaporated plasma in the heated chromosphere on the left-sided footpoint. What we want to stress here is that the right-sided footpoint also contributes to the plasma accumulation in the corona that finally leads to filament formation. When the evaporated plasma rushes into the corona from the left-sided footpoint, a shock wave is excited ahead, as shown in the right panel of Figure \ref{fig2}. When the shock reaches the right-sided footpoint, it heats the chromosphere, leading to chromospheric evaporation  at the right-sided footpoint as well. The shock wave might bounce back and forth between the two footpoints before fading away. The effect of the shock wave is two-fold: On the one hand, it heats the chromosphere, leading to more plasma evaporating into the corona; On the other hand, the leftward-propagating shock wave suppresses the rightward moving plasma from draining down to the right-sided footpoint, allowing more plasma accumulating in the corona so as to form a filament thread. Once a small segment of filament thread is formed, it will grow slowly via siphon flows from both footpoints even no extra heating is imposed at the footpoints, as demonstrated by \citet{xia11}.

To summarize, in this paper we performed 1D hydrodynamic simulations including gravity, thermal conduction, radiative cooling, and heating, with the purpose to unify the evaporation--condensation model and the direct injection model for filament formation. It is demonstrated that when the localized heating is situated in the upper chromosphere, the local plasma is heated to evaporate into the corona. At a certain stage, the density-enhanced corona cools down to form a filament thread via thermal instability; When the localized heating is situated in the lower chromosphere, the local plasma is heated, and its enhanced gas pressure pushes the cold upper chromospheric material to be injected into the corona like surges. Depending on the depth of the magnetic dip and the injection speed, the cold material might be trapped in the coronal magnetic dip, forming a quasi-static filament, or keep moving along the magnetic loop until draining down to the other footpoint of the magnetic loop.

It is noted that, following previous 1D hydrodynamic simulations, the localized chromospheric heating in our work is realized by depositing energy in situ. In reality, such heating is believed to result from low atmosphere magnetic reconnection. Such a process can be reproduced only via 2- or 3-dimensional magnetohydrodynamic simulations, which are devoted for future research.

\acknowledgments
This research was supported by the Chinese foundations NSFC (11961131002) and National Key Research and Development Program of China (2020YFC2201201).

\end{document}